\documentclass [preprint, prd, aps, nofootinbib]{revtex4}

\usepackage{graphicx}

\begin{document}
\draft
\title{\large \bf From spacetime foam to holographic foam cosmology}

\author{\bf Michele Arzano$^{(1)}$ \footnote{arzano@physics.unc.edu},
Thomas W. Kephart$^{(2)}$ \footnote{
thomas.w.kephart@vanderbilt.edu} 
and Y. Jack Ng$^{(1)}$ \footnote{yjng@physics.unc.edu 
(corresponding author)}}
\address{(1) Institute of Field Physics, Department
of Physics and Astronomy, University of North Carolina, Chapel
Hill, NC 27599, USA\\
(2) Department of Physics and Astronomy, Vanderbilt University,
Nashville, TN 37235, USA}

\bigskip

\begin{abstract}

Due to quantum fluctuations, spacetime is foamy on small scales.  For 
maximum spatial resolution of the geometry of spacetime, the
holographic model of spacetime foam stipulates that the uncertainty
or fluctuation of distance $l$ is given, on the average,
by $(l l_P^2)^{1/3}$ where $l_P$ is
the Planck length.   Applied to cosmology, it predicts that
the cosmic energy is of critical density and the cosmic entropy is the
maximum allowed by the holographic principle.  In addition, it requires
the existence of unconventional (dark) energy/matter and 
accelerating cosmic expansion in the present era.  We will argue that a 
holographic foam cosmology of this type has the potential to become a 
full fledged competitor (with distinct testable consequences) for scalar 
driven inflation.

\vspace{3cm}

{\it This paper is dedicated to Rafael Sorkin to celebrate his 
contributions to physics; to appear on the Sorkin60 website.}



\end{abstract}
\maketitle




Before last century, spacetime was regarded as nothing more than a passive and
static arena in which events took place.  Early last century, Einstein's 
general relativity changed that viewpoint and promoted spacetime to an
active and dynamical entity.  Nowadays, following
Wheeler\cite{Wheeler}, some of us also believe that space is composed of an
ever-changing geometry and topology called spacetime foam, and that the
foaminess is due to quantum fluctuations \cite{Ford} of spacetime.  In this paper, we will 
consider the holographic model of spacetime foam, and apply it to the cosmos
of the present era. 

The magnitude of spacetime fluctuations can be quantified by 
the average uncertainty $\delta l$ in
the measurement of distance $l$.  In principle, $\delta l$ can depend on
both $l$ and the Planck length $l_P \equiv (\hbar G/c^3)^{1/2} \sim
10^{-33}$ cm, and hence can be written as $\delta l \gtrsim
l^{1 - \alpha} l_P^{\alpha}$, with $\alpha \sim 1$ parametrizing the
spacetime foam models.
We begin by recalling a simple argument \cite{llo04,gio04}
for the choice of $\alpha = 2/3$.  Consider mapping the geometry of
spacetime for a volume of spatial extent $l$ and temporal extent $l/c$.  One
way is to fill space with clocks, exchanging signals with each other and
measuring the signals' times of arrival.  This process is a kind of
computation and is hence constrained by the
Margolus-Levitin theorem\cite{mar98} in
quantum computation, that bounds
the rate of elementary logical operations that can be
performed within the volume by the amount of energy that is available for
computation, which is $Mc^2$ for the present setup with $M$ being the total
mass of the clocks, divided by Planck's constant.  To avoid black hole
formation, $M$ must be less than $l/G$, corresponding to an energy density
$\rho \lesssim (l l_P)^{-2}$.  (Here and henceforth we neglect
multiplicative constants of order unity, and set $c=1=\hbar$.)  It follows
that the number of elementary operations or events that can occur in this
spacetime volume is bounded by $l^2/l_P^2$.  In other words, if one regards
the elementary events partitioning the spacetime volume into ``cells'', then
the number of cells is bounded by the 
surface area of the spatial volume, and each cell
occupies a spacetime volume of $l^4 / (l^2 / l_P^2) = l^2 l_P^2$ on the
average.

The maximum
spatial resolution of the geometry is obtained if each clock ticks once
in time $l$.  Then each cell occupies a spatial volume of $l^2 l_P^2 / l
= l l_P^2$, yielding an average separation between neighboring cells of
$l^{1/3} l_P^{2/3}$.  This spatial separation can now be interpreted as
the average minimum uncertainty $\delta l$ in the measurement of
a distance $l$, i.e., $\delta l \gtrsim l^{1/3} l_P^{2/3}$.

Two remarks are in order.  First, in the above argument, maximal 
spatial resolution is
possible only if the maximum energy density $\rho \sim (l l_P)^{-2}$
is available (to map the geometry
of the spacetime region) without causing a gravitational collapse.
Secondly, 
since, on the average, each cell occupies a spatial volume of $l l_P^2$,
a spatial region of size $l$ can contain no more than $l^3/(l l_P^2) = 
(l/l_P)^2$ cells. 
Hence, the above argument also
shows that this spacetime foam model (yielding the maxiumum spatial
resolution of the geometry of spacetime) corresponds to the case of
maximum number of bits of information $l^2 /l_P^2$
in a spatial region of size $l$, that is
allowed by the holographic principle\citep{tho93}, acording to which,
the maximum amount of information stored in a region of space scales as
the area of its two-dimensional surface, like a hologram.  For good
reason, this spacetime foam model\cite{ng94} has come 
to be known, in recent years, as the holographic model\cite{ng01}.  (We 
have just illustrated how a holographic description arises from local 
small scale spacetime foam physics!)  Alternatively, the holographic 
principle can also be derived by using the above argument as follows:
Consider a spatial region of size $l$ containing particles each of which
carries one bit of information.  Heisenberg's uncertainty principle
dictates that each particle/bit has a momentum greater than $l^{-1}$.
Now matter can embody the maximum amount of information when it is 
converted to energetic, massless particles.  In that case, each 
particle/bit has an energy no less than $l^{-1}$.  But, as shown above, 
the maximum amount of energy inside the spatial region is bounded by
$l^3 \times (l l_P)^{-2} = l l_P^{-2}$.  Hence the maximum number of
bits in a spatial region of size $l$ is bounded by $l l_P^{-2} / l^{-1}
= (l/l_P)^2$.

So far, we have confined our attention to a static spacetime region with
low spatial curvature.  The whole discussion can be generalized to the 
case of an expanding 
universe by the substitution of $l$ by $H^{-1}$ in the expressions for 
energy and entropy densities, where $H$ is the Hubble parameter.  Thus,
applied to cosmology, the holographic model of spacetime foam  
predicts that (1) the cosmic energy density is critical 
$\rho \sim (H/l_P)^2$, and (2) the universe
of Hubble size $R_H$ contains $\sim H R_H^3/ l_P^2$ bits of information.  We 
call this cosmology the holographic foam cosmology (HFC).  (For earlier 
discussions of holographic cosmology, see \cite{FandS}.)

It is instructive to compare the holographic model we have discussed
above in the mapping of the geometry of spacetime
with the one that corresponds to spreading the spacetime cells uniformly
in both space and time.  For the latter case, each cell has
the size of $(l^2 l_P^2)^{1/4} =
l^{1/2} l_P^{1/2}$ both spatially and temporally, i.e., each clock ticks
once in the time it takes to communicate with a neighboring clock.  Since
the dependence on $l^{1/2}$ is the hallmark of a random-walk fluctuation,
this spacetime foam model corresponding to  $\delta l \gtrsim
(l l_P)^{1/2}$ is called the random-walk model\cite{dio89}.  
Compared to the holographic model, the random-walk model predicts a 
coarser spatial resolution, i.e., a larger distance fluctuation, 
in the mapping of spacetime geometry.  It 
also yields a smaller bound on the information content in a spatial 
region, viz., $(l/l_p)^2 / (l/l_P)^{1/2} = (l/l_P)^{3/2}$.
We further note that the spatial resolution for the
random-walk model, unlike the holographic model,
does not require the maximum total
energy because the clocks can tick less frequently than once in
the amount of time $l^{1/2} l_P^{1/2}$ (so long as each clock
ticks at least once in the entire time duration of $l$.)

We are now in a position to draw a useful conclusion\cite{cnvd,llo04}
from the {\it observed} cosmic critical density in the present era
(consistent with the prediction of the HFC)
$\rho \sim H_0^2/G \sim (R_H l_P)^{-2}$ (about $10^{-9}$
joule per cubic meter),
where $H_0$ is the present Hubble parameter of the observable universe.
Treating the whole universe as a computer\cite{llo02, llo04}, one can
apply the Margolus-Levitin theorem to conclude that the universe
computes at a rate $\nu$ up to $\rho R_H^3 \sim R_H l_P^{-2}$
($\sim 10^{106}$ op/sec), for a total of $(R_H/l_P)^2$
($\sim10^{122}$) operations during its lifetime so far.
If all the information of this huge computer is stored in ordinary
matter, then we can apply standard methods of statistical mechanics 
to find that the total number $I$ of bits is $[(R_H/l_P)^2]^{3/4} =
(R_H/l_P)^{3/2}$ ($\sim 10^{92}$).
It follows that each bit flips once in the amount of time given by
$I/\nu \sim (R_H l_P)^{1/2}$ ($\sim 10^{-14}$ sec).  On
the other hand, the average separation of neighboring bits is
$(R_H^3/I)^{1/3} \sim (R_H l_P)^{1/2}$ ($\sim 10^{-3}$
cm).  Hence, assuming only ordinary matter exists to store all the
information in the universe results in the conclusion that the time
to communicate with neighboring bits is equal to the time for each
bit to flip once.  It follows that the accuracy to which ordinary
matter maps out the geometry of spacetime corresponds exactly to
the case of events
spread out uniformly in space and time discussed above for the case
of the random-walk model of spacetime foam.

Interestingly, it has been shown \cite{cnvd} 
that the sharp images of distant quasars or active galactic nuclei 
observed at the Hubble Space
Telescope have ruled out the random-walk model.  More specifically,
the presence of an Airy ring in the image of the quasar-like object
PKS1413+135 \cite{Perlman}
indicates that the amount of light scattering that could be
caused by spacetime foam is much less than that predicted by the 
random-walk model.  From the demise of the 
random-walk model and the fact that ordinary matter only contains an
amount of information dense enough to map out spacetime at a level
consistent with the random-walk model, one can now infer that
spacetime must be mapped to a finer spatial accuracy than that which
is possible with the use of ordinary matter.  The natural 
conclusion \cite{cnvd} one draws is that unconventional 
(dark\cite{Turner}) energy/matter exists!
Note that this argument does not make use of
the evidence from recent cosmological (supernovae, cosmic microwave
background, and galaxy clusters)
observations.  On the other hand,
the demise of the random-walk model and the existence of dark
energy/matter are consistent with but do not necessarily mean
the vindication of the holographic model.  Fortunately, in the next
few years, the Very Large Telescope Interferometers in Chile could
test \cite{cnvd}
the holographic model by observing more distant quasars with
their large apertures and long baselines.

But for now, the fact that our
universe is observed to be at or very close to its critical density 
must be taken as solid albeit indirect evidence in favor of the 
holographic model\cite{cnvd} because, as discussed above, it is the only
model that requires, for its consistency,
the maximum energy density without causing gravitational collapse. 
Henceforth we will concentrate on the holographic model.  What can be
said about the unconventional (dark) energy?  According
to the HFC, the cosmic density is
$\rho \sim (H/l_P)^2 \sim (R_H l_P)^{-2}$ and the 
cosmic entropy is given by $I \sim H R_H^3/l_P^2
\sim (R_H /l_P)^2$.  Hence the average energy carried by each bit is
$\rho R_H^3/I \sim R_H^{-1}$ ($\sim 10^{-31}$ eV).  Such
long-wavelength \cite{JK} bits or ``particles'' carry negligible kinetic energy.
Since pressure (energy density) is given by kinetic energy minus (plus) 
potential energy, a negligible kinetic energy means that
the pressure of the unconventional energy is roughly equal to minus its 
energy density, leading to accelerating cosmic
expansion.  This scenario is very similar to that of quintessence
\cite{RatraPeebles}, but it has its origin in 
local small scale physics --- specifically, the holographic spacetime 
foam.

According to the HFC, it takes each unconventional bit
the amount of time
$I/\nu \sim R_H$ to flip.  Thus, on the average, each bit flips
once over the course of cosmic history.  Compared to the conventional
bits carried by ordinary matter, these bits are rather passive and
inert.  This is understandable since each unconventional bit has,
at its disposal, only
such a minuscule amount of energy.  But together they supply the
missing mass of the universe.  Accelerating the cosmic expansion is
a relatively simple task, computationally speaking.  It is also
interesting to note \cite{llo04}
that if the universe contains only ordinary
matter, then computationally it is a supreme parallel computer with
its subregions working almost independently, as each bit flips once in
the same amount of time it takes to communicate with its neighboring
bits.  But if the universe contains only unconventioanl energy
which encodes the maximum number of bits allowed by the holographic
principle, then it can be regarded as a supreme serial computer which
operates as a single unit, as each bit flips once in the amount of
time it takes for a light signal to cross the whole universe.  The
universe actually contains both types of matter/energy,
though apparently with
vastly more bits stored in dark energy/matter than in ordinary matter.

Due to the enormous number of the unconventional bits, neighboring
bits are separated from each other, on the average, by only
$(R_H^3/I)^{1/3} \sim R_H^{1/3} l_P^{2/3}$.   But with a wavelength
comparable to the Hubble radius, these bits/particles practically
sit on top of one another (similar to overlapping wave functions in
superfluid or superconductor with large coherent lengths),
leading to a rather uniform distribution of energy density 
which, we
recall, is given by the geometric mean of the infrared and
ultraviolet energy scales $\rho \sim (H/l_P)^2
\sim (R_H l_P)^{-2}$.  In this
regard, the unconventional energy plays a role akin to that of an
effective cosmological ``constant'' $\Lambda \sim R_H^{-2}$.
Such an effective cosmological constant \cite{interpretation}
was shown to arise in theories like unimodular gravity by Sorkin and 
others \cite{Sorkin}, and in
recent years, in other different contexts as well \cite{Cohen,Thomas}.
Thomas \cite{Thomas} has also argued that, in an expanding 
universe, holographic contributions to the cosmological constant
are at most of the same order as the energy density of the
dominant matter component, thus ameliorating the coincidence
problem, and, at the same time, providing a technically natural
solution to the cosmological constant problem.


In this paper we 
have only applied the HFC to the present cosmic era, predicting 
an accelerating expansion.  Whether the HFC can
explain the early universe has yet to be seen.  Here we conclude on
an optimistic note with the following observations.  One of the HFC's
main features, that the cosmic energy is of critical density, is 
a hallmark of 
the inflationary universe paradigm.  Thus the flatness problem is 
automatically solved.  Another requirement on the HFC is for the
model to provide sufficient density perturbations to account for
the observed structure in the universe.  We beleive this is 
possible, as the model already contains the essence of a 
k-essence model.  Turning to the horizon problem, we note that
spacetime foam physics is, in a nutshell, black hole physics in
a quantum setting, hence it is intimately related to wormhole 
physics.   But it has been argued in \cite{HandK} that wormholes
in a Friedmann-Robertson-Walker universe can be used to solve the
horizon problem.  Thus it is possible that the HFC can meet its 
many challenges, solving naturally the classical
cosmological problems and predicting new phenomena.  Further work 
along these lines is warranted.

\newpage


M.A. was supported by a Dissertation Completion Research Fellowship funded
by the University of North Carolina Graduate School;
T.W.K  was supported in part by U.S. DOE grant $\#$ DE-FG05-85ER40226;
Y.J.N was supported in part by DOE grant $\#$ DE-FG02-06ER41418 and
the Bahnson Fund at the University of North Carolina.

\vspace{2cm}

{\it Rafael Sorkin is an extraordinary physicist, admired for his 
insights and originality.  He is also a 
gentle and wonderful man.  Happily and humbly
Y.J.N. dedicates this paper to him to celebrate his remarkable
contributions to physics.}


\begin{references}
\bibitem{interpretation}
The long-wavelength
bits/particles can be interpreted as constituting the degrees
of freedom for the cosmological ``constant'', just as tiny strings
are those of black holes in string theory.
\end{references}
\end{document}